\documentclass[12pt]{article}
\usepackage{graphicx}
\usepackage[cp1251]{inputenc}
\usepackage{rotating}
\begin{document}
 \noindent {\footnotesize\it Astronomy Letters, 2018, Vol. 44, No 3, pp. 184--192.}
 \newcommand{\dif}{\textrm{d}}

 \noindent
 \begin{tabular}{llllllllllllllllllllllllllllllllllllllllllllll}
 & & & & & & & & & & & & & & & & & & & & & & & & & & & & & & & & & & & & & \\\hline\hline
 \end{tabular}

  \vskip 0.5cm
 \centerline{\bf\Large Testing the distance scale of the Gaia~TGAS catalogue }
 \centerline{\bf\Large by kinematic method}
 \bigskip
 \bigskip
  \centerline
 {
 V.V. Bobylev\footnote [1]{e-mail: vbobylev@gao.spb.ru} and
 A.T. Bajkova
 }
 \bigskip

  \centerline{\small\it
 Pulkovo Astronomical Observatory, Russian Academy of Sciences,}

  \centerline{\small\it
 Pulkovskoe sh. 65, St. Petersburg, 196140 Russia}
 \bigskip
 \bigskip
 \bigskip

 {
{\bf Abstract}---We have studied the simultaneous and separate
solutions of the basic kinematic equations obtained using the
stellar velocities calculated on the basis of data from the Gaia
TGAS and RAVE5 catalogues. By comparing the values of
$\Omega_0^{'}$ found by separately analyzing only the
line-of-sight velocities of stars and only their proper motions,
we have determined the distance scale correction factor p to be
close to unity, $0.97\pm0.04$. Based on the proper motions of
stars from the Gaia TGAS catalogue with relative trigonometric
parallax errors less than 10\% (they are at a mean distance of 226
pc), we have found the components of the group velocity vector for
the sample stars relative to the Sun
$(U,V,W)_\odot=(9.28,20.35,7.36)\pm(0.05,0.07,0.05)$ km s$^{-1}$,
the angular velocity of Galactic rotation $\Omega_0=27.24\pm0.30$
km s$^{-1}$ kpc$^{-1}$, and its first derivative
$\Omega_0^{'}=-3.77\pm0.06$ km s$^{-1}$ kpc$^{-2}$; here, the
circular rotation velocity of the Sun around the Galactic center
is $V_0=218\pm6$ km s$^{-1}$ kpc (for the adopted distance
$R_0=8.0\pm0.2$ kpc), while the Oort constants are
$A=15.07\pm0.25$ km s$^{-1}$ kpc$^{-1}$ and $B=-12.17\pm0.39$ km
s$^{-1}$ kpc$^{-1}$, $p=0.98\pm0.08$. The kinematics of Gaia TGAS
stars with parallax errors more than 10\% has been studied by
invoking the distances from a paper by Astraatmadja and
Bailer-Jones that were corrected for the Lutz–Kelker bias. We show
that the second derivative of the angular velocity of Galactic
rotation $\Omega_0^{''}=0.864\pm0.021$ km s$^{-1}$ kpc$^{-3}$ is
well determined from stars at a mean distance of 537 pc. On the
whole, we have found that the distances of stars from the Gaia
TGAS catalogue calculated using their trigonometric parallaxes do
not require any additional correction factor.
  }

\medskip
DOI: 10.1134/S1063773718020020

 \subsection*{INTRODUCTION}
The high expectations of specialists in studying the structure and
kinematics of the Galaxy are associated with the determination of
highly accurate trigonometric parallaxes and proper motions for
hundreds of millions of stars whose observations are planned in
the Gaia project (Prusti et al. 2016). The first results were
already published in 2016. In particular, the proper motions of
stars were found by comparing their positions measured from the
Gaia satellite with those from the Hipparcos/Tycho (1997)
catalogue with an epoch difference of about 24 years. This version
is designated as TGAS (Tycho--Gaia Astrometric Solution, Brown et
al. 2016; Lindegren et al. 2016) and contains the trigonometric
parallaxes and proper motions of $\sim$2 million stars. The proper
motions of $\sim$90 000 stars common to the Hipparcos catalogue
were measured with a mean error of $\sim$0.06 mas yr$^{-1}$, while
for the remaining stars this error is $\sim$1 mas yr$^{-1}$ (Brown
et al. 2016).

The mean random measurement error of the trigonometric parallaxes
in the TGAS catalogue is $\sim$0.3 mas, to which a systematic
component of $\sim$0.3 mas should be added (Brown et al. 2016).
Having analyzed the kinematic characteristics of 19 open star
clusters, van Leeuwen et al. (2017) concluded that the parallaxes
in this catalogue have a good quality. It should be particularly
noted that the new distance to the Pleiades, $134\pm4$~pc, agrees
better with its known reliable determinations, $133.7\pm1.2$~pc
(An et al. 2007; Kim et al. 2016), in particular, with its VLBI
measurements, $136.2\pm1.2$~pc (Melis et al. 2014), than with its
value in the Hipparcos catalogue, $120\pm2$~pc (van Leeuwen 2009).
Fortunately, the Pleiades distance problem turned out to be local
in the Hipparcos catalogue.

A comparison of the distances to Cepheids and RR Lyr variables
from the TGAS catalogue with their distances estimated by other
methods (for example, based on the period--luminosity relation or
from Hubble Space telescope astrometry) has shown excellent
agreement between the results up to distances of $\sim$2 kpc
(Casertano et al. 2017; Benedict et al. 2017; Clementini et al.
2017). Based on the kinematic method, Bobylev and Bajkova (2017)
showed that the distances to the stars from a sample of OB stars
(within $\sim$3--4 kpc of the Sun) do not require the introduction
of any correction factors.

However, there are reports of several authors about the detection
of a systematic offset in the stellar parallaxes from the TGAS
catalogue. In particular, Stassun and Torres (2016) detected such
an offset, $-0.25\pm0.05$~mas, with respect to 158 calibration
eclipsing binary stars. Having analyzed nearby stars (within 25
pc), Jao et al. (2017) found that the TGAS parallaxes are, on
average, $0.24\pm0.02$~mas smaller than the trigonometric
parallaxes for these stars measured from the ground. Applying the
statistical method revealed local anomalies in the distances when
analyzing the space velocities of stars calculated from a
combination of RAVE--TGAS and LAMOST--TGAS data (Sch\"onrich and
Aumer 2017). The RAVE (RAdial Velocity Experiment, Steinmetz et
al. 2006) and LAMOST (Large sky Area Multi-Object Fiber
Spectroscopic Telescope, Luo et al. 2015) catalogues contain the
results of large-scale line-of-sight velocity measurements for
stars predominantly in the southern and northern hemispheres of
the celestial sphere, respectively. Thus, we see that the
properties of the TGAS distance scale have not yet been completely
studied and, therefore, an analysis of this distance scale with
the application of independent approaches is topical.

The goal of this paper is to study the TGAS stellar velocity field
based on the separate solutions of the basic kinematic equations
from the line-of-sight velocities and proper motions of stars.
Since the line-of-sight velocities and proper motions of stars are
determined by fundamentally different methods, our separate
solutions of the basic kinematic equations allow the consistency
between these data to be investigated from a kinematic viewpoint.

 \subsection*{METHODS}
We know three stellar velocity components from observations: the
line-of-sight velocity $V_r$ and the two tangential velocity
components $V_l=4.74r\mu_l\cos b$ and $V_b=4.74r\mu_b$ along the
Galactic longitude $l$ and latitude $b,$ respectively, expressed
in km s$^{-1}$. Here, the coefficient 4.74 is the ratio of the
number of kilometers in an astronomical unit to the number of
seconds in a tropical year, and $r$ is the stellar heliocentric
distance in kpc. The proper motion components $\mu_l\cos b$ and
$\mu_b$ are expressed in mas yr$^{-1}$.

To determine the parameters of the Galactic rotation curve, we use
the equations derived from Bottlinger’s formulas, in which the
angular velocity $\Omega_0$ is expanded into a series to terms of
the second order of smallness in $r/R_0:$
\begin{equation}
 \begin{array}{lll}
 V_r&=&-U_\odot\cos b\cos l-V_\odot\cos b\sin l-W_\odot\sin b\\
 &&+R_0(R-R_0)\sin l\cos b\Omega^{'}_0
 +0.5R_0(R-R_0)^2\sin l\cos b\Omega^{''}_0,
 \label{EQ-1}
 \end{array}
 \end{equation}
 \begin{equation}
 \begin{array}{lll}
 V_l&=& U_\odot\sin l-V_\odot\cos l-r\Omega_0\cos b\\
    && +(R-R_0)(R_0\cos l-r\cos b)\Omega^{'}_0
    +0.5(R-R_0)^2(R_0\cos l-r\cos b)\Omega^{''}_0,
 \label{EQ-2}
 \end{array}
 \end{equation}
 \begin{equation}
 \begin{array}{lll}
 V_b&=&U_\odot\cos l\sin b + V_\odot\sin l \sin b-W_\odot\cos b\\
    && -R_0(R-R_0)\sin l\sin b\Omega^{'}_0
    -0.5R_0(R-R_0)^2\sin l\sin b\Omega^{''}_0,
 \label{EQ-3}
 \end{array}
 \end{equation}
Here, $R$ is the distance from the star to the Galactic rotation
axis:
  \begin{equation}
 R^2=r^2\cos^2 b-2R_0 r\cos b\cos l+R^2_0.
 \end{equation}
The quantity $\Omega_0$ is the angular velocity of Galactic
rotation at the solar distance $R_0,$ $\Omega_0^{'}$ is its first
derivative, $V_0=|R_0\Omega_0|;$ the Oort constants $A$ and $B$
can be found from the expressions
\begin{equation}
  A=-0.5\Omega_0^{'}R_0,\quad B=-\Omega_0+A, \label{AB}
\end{equation}
written in such a way that the following relations hold:
$A-B=\Omega_0$ and $A+B=-(\Omega_0+\Omega_0^{'}R_0)$. In this
paper we adopt $R_0=8.0\pm0.2$~kpc that Vall\'ee (2017) found in
his recent review as the most probable value.

When simultaneously solving the system of conditional equations
(1)–(3) by the least-squares method, we find and six unknowns
$U_\odot,V_\odot,W_\odot,\Omega_0,\Omega_0^{'},\Omega_0^{''}$. We
can find five unknowns
$U_\odot,V_\odot,W_\odot,\Omega_0^{'},\Omega_0^{''}.$ from Eq. (1)
by analyzing only the line-of-sight velocities. When analyzing
only the proper motions, it is convenient to solve the system of
equations (2)--(3), which allows all six unknowns to be
determined.

 \subsection*{DATA}
Our sample includes stars from the RAVE5 catalogue (Kunder et al.
2017) with measured line-of-sight velocities and with estimated
trigonometric parallaxes and proper motions from the Gaia DR1
catalogue. Note that the RAVE5 catalogue contains more than 60 000
stars for which the line-of-sight velocities were measured several
times. Therefore, we ultimately produced a sample in which each
star is presented once. In the case where a star has several
line-of-sight measurements, we did not perform any averaging but
took the measurement with the smallest line-of-sight velocity
error. This sample includes a total of $\sim$200 000 stars.

There are stars with large line-of-sight velocities $|V_r|>600$ km
s$^{-1}$ in the RAVE5 catalogue. Such values were typically
obtained from low-quality spectra, with a small signal-to-noise
ratio. Therefore, we do not use the stars with such velocities,
nor do we use the stars with large random errors in the
line-of-sight velocity $\sigma_{V_r}$. As a result, for the
selection of candidates without significant random observational
errors we took the stars that satisfied the following criteria:
\begin{equation}
 \begin{array}{ccc}
        |V_r|<600~{\hbox {\rm km s$^{-1}$}},\qquad
 \sigma_{V_r}<5~{\hbox {\rm km s$^{-1}$}},\\
 |\mu_\alpha\cos\delta|<400~{\hbox {\rm mas yr$^{-1}$}},\qquad
           |\mu_\delta|<400~{\hbox {\rm mas yr$^{-1}$}},\\
 \sqrt{U^2+V^2+W^2}<200~{\hbox {\rm km s$^{-1}$}},
 \label{cut-1}
 \end{array}
 \end{equation}
where the velocities $U,V,$ and $W$ were freed from the Galactic
differential rotation, i.e., they are the residual ones. Any known
Galactic rotation curve, for example, from Bobylev and Bajkova
(2017) or Rastorguev et al. (2017), is suitable to perform this
procedure.

 \subsection*{RESULTS}
Table 1 gives the kinematic parameters found from stars with
measured proper motions and trigonometric parallaxes from the Gaia
DR1 catalogue. The parameters in this table were calculated at
relative trigonometric parallax errors $\sigma_\pi/\pi<10\%.$ We
used a total of 53 173 stars. At such small radii of the sample
the second derivative of the angular velocity of Galactic rotation
is determined very poorly. Therefore, the table provides the
results of the solution of Eqs. (1)--(3) without $\Omega^{''}$. At
small $\sigma_\pi/\pi$ the influence of the Lutz–Kelker (1973)
bias is negligible, while at large $\sigma_\pi/\pi$ this bias
should be taken into account (Astraatmadja and Bailer-Jones 2016a,
2016b). The first, second, and third columns in the table present
the results of the simultaneous solution, those obtained only from
the line-of-sight velocities, and those obtained only from the
proper motions, respectively. Table 1 also gives the error per
unit weight $\sigma_0$ determined when solving the conditional
equations (1)--(3) by the least-squares method, which is close in
its meaning to the residual velocity dispersion for the sample of
stars being analyzed averaged over all directions. The number of
stars used $N_\star$ is specified, the Oort constants $A$ and $B$
calculated from Eqs. (5) are given.

 \begin{table}[t]
 \caption[]
  {\small
The Galactic rotation parameters found from stars with measured
line-of-sight velocities and proper motions provided that the
relative trigonometric parallax errors $\sigma_\pi/\pi$ are less
than 10\%
  }
  \begin{center}  \label{t:01}
  \small
  \begin{tabular}{|l|r|r|r|r|r|}
   \hline
   Parameters                   &  $V_r,V_l,V_b$  &          $V_r$  &      $V_l,V_b$  \\\hline
   $U_\odot,$    km s$^{-1}$    & $ 9.15\pm0.18$  & $ 9.59\pm0.29$  & $ 8.91\pm0.22$  \\
   $V_\odot,$    km s$^{-1}$    & $20.41\pm0.15$  & $19.94\pm0.27$  & $20.55\pm0.19$  \\
   $W_\odot,$    km s$^{-1}$    & $ 7.74\pm0.12$  & $ 8.26\pm0.18$  & $ 7.51\pm0.17$  \\
 $\Omega_0,$     km s$^{-1}$ kpc$^{-1}$ & $ 27.6\pm1.4$   &            ---  & $27.9 \pm1.7$   \\
 $\Omega_0^{'},$ km s$^{-1}$ kpc$^{-2}$ & $-3.89\pm0.23$  & $-3.70\pm0.38$  & $-3.83\pm0.29$  \\
   $\sigma_0,$   km s$^{-1}$    &          27.51  &          26.54  &          27.98  \\
               $N_\star$        &          53173  &          53173  &          53173  \\
    $A,$ km s$^{-1}$ kpc$^{-1}$ & $ 15.56\pm0.91$ & $ 14.81\pm1.52$ & $ 15.33\pm1.15$ \\
    $B,$ km s$^{-1}$ kpc$^{-1}$ & $-12.06\pm1.68$ &             --- & $-12.62\pm2.04$ \\
 \hline
 \end{tabular}\end{center} \end{table}
 \begin{table}[t]
 \caption[]
  {\small
The Galactic rotation parameters found only from the proper
motions of stars $(V_l,V_b)$ from the Gaia TGAS catalogue with
relative trigonometric parallax errors $\sigma_\pi/\pi<10\%$
  }
  \begin{center}  \label{t:02}
  \small
  \begin{tabular}{|l|r|r|r|r|r|}
   \hline
   Parameters                   &         I       &           II    &        III      &      IV  \\\hline
   $U_\odot,$    km s$^{-1}$    &  $ 9.25\pm0.06$ &  $ 9.46\pm0.09$ &  $ 9.59\pm0.17$ &  $ 8.75\pm0.16$ \\
   $V_\odot,$    km s$^{-1}$    &  $16.64\pm0.09$ &  $23.56\pm0.13$ &  $23.31\pm0.25$ &  $21.71\pm0.23$ \\
   $W_\odot,$    km s$^{-1}$    &  $ 7.23\pm0.06$ &  $ 7.49\pm0.09$ &  $ 7.45\pm0.16$ &  $ 7.42\pm0.15$ \\
 $\Omega_0,$     km s$^{-1}$ kpc$^{-1}$ &  $26.42\pm0.38$ &  $27.78\pm0.68$ &  $27.21\pm0.87$ &  $28.41\pm0.99$ \\
 $\Omega_0^{'},$ km s$^{-1}$ kpc$^{-2}$ &  $-3.71\pm0.08$ &  $-3.97\pm0.14$ &  $-3.67\pm0.18$ &  $-4.09\pm0.21$ \\
   $\sigma_0,$   km s$^{-1}$    &           23.18 &           27.94 &          28.17  &           28.35 \\
   $A,$ km s$^{-1}$ kpc$^{-1}$  & $ 14.83\pm0.32$ & $ 15.86\pm0.57$ & $ 14.66\pm0.71$ & $ 16.37\pm0.83$ \\
   $B,$ km s$^{-1}$ kpc$^{-1}$  & $-11.59\pm0.50$ & $-11.91\pm0.89$ & $-12.55\pm1.12$ & $-12.04\pm1.29$ \\
             $\overline r,$ pc  &             233 &             195 &            283  &            229  \\
               $N_\star$        &          202360 &          130472 &          44476  &          46943  \\
               $p$              &   $1.00\pm0.08$ &   $0.93\pm0.07$ &   $1.01\pm0.05$ &   $0.91\pm0.04$ \\
 \hline
 \end{tabular}\end{center} \end{table}
 \begin{figure} {\begin{center}
 \includegraphics[width=80mm]{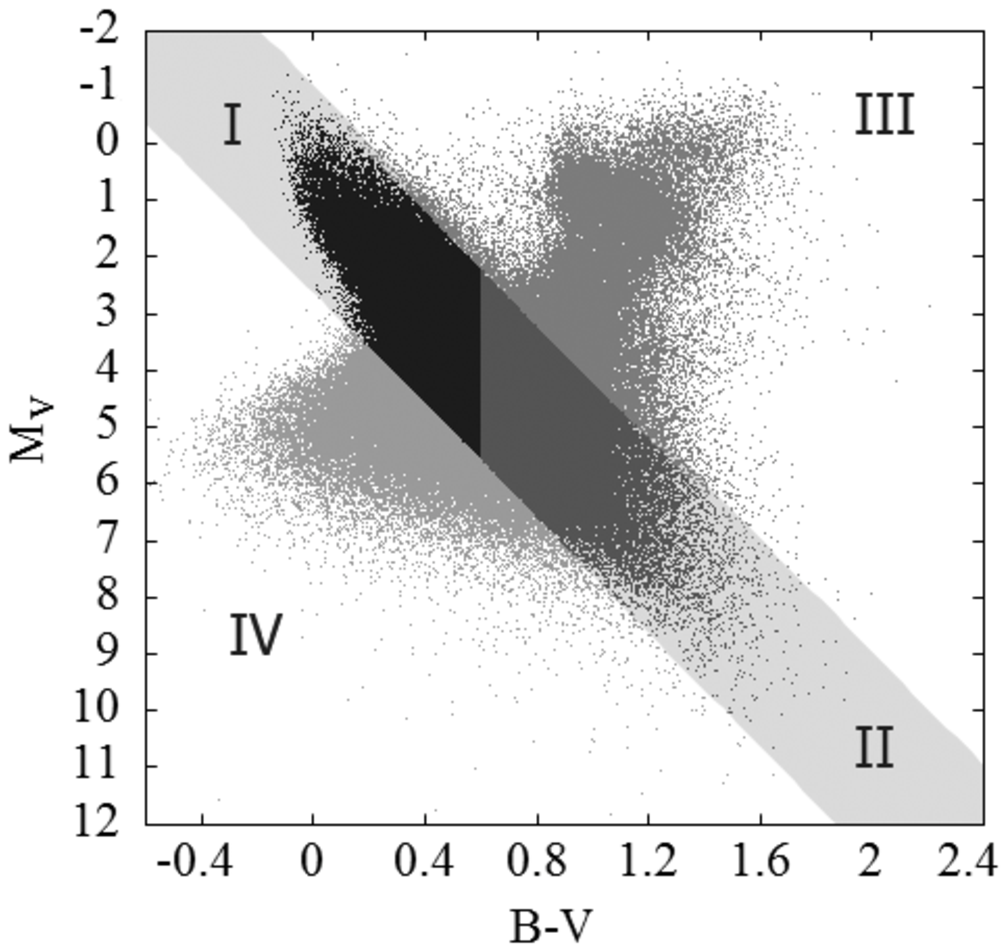}
 \caption{
Distribution of the four regions on the Hertzsprung--Russell
diagram constructed for stars from the Gaia TGAS catalogue with
relative trigonometric parallax errors less than 10\%.
 }
  \end{center} } \end{figure}

The values of $\Omega_0^{'}$ obtained in the separate solutions
are of interest for checking the distance scale used. This method
is based on the fact that the line-of-sight velocity errors do not
depend on the distance errors, while the errors in the tangential
components depend on the latter. Therefore, a comparison of the
values of $\Omega_0^{'}$ found by various methods allows the
distance scale correction factor $p$ to be determined
(Zabolotskikh et al. 2002; Rastorguev et al. 2017). According to
the data in Table 1, this correction factor is
$p=(-3.70)/(-3.83)=0.97\pm0.04;$ its error was calculated from the
relation
 $$
 \sigma_p=(\sigma_{\Omega'_{0r}}/\Omega'_{0t})-
          (\Omega'_{0r}\sigma_{\Omega'_{0t}}/\Omega'^2_{0t}),
 $$
where the quantity $\Omega'_{0(l,b)}$ is denoted by
$\Omega'_{0t}$. Note that Bobylev and Bajkova (2017) found
$p=0.96$ from the kinematics of distant OB stars. Having analyzed
the line-of-sight velocities and proper motions of more nearby red
giants from the RAVE5 catalogue, Vityazev et al. (2017) found this
correction factor to be 0.96, but its value from main-sequence
dwarfs was 0.72.

Table 2 gives the Galactic rotation parameters found from the
proper motions of all stars (from both southern and northern
hemispheres of the celestial sphere) from the Gaia TGAS catalogue
with relative parallax errors $\sigma_\pi/\pi<10\%$. We divided
all stars into four groups (I, II, III, and IV) according to their
positions on the Hertzsprung--Russell diagram shown in the figure.
To construct this diagram, we used Tyho-2 photometric data, $B_T$
and $V_T$, by taking into account the well-known calibration
relations $B-V=0.85(B_T-V_T)$ and $V=V_T-0.09(B_T-V_T)$ to pass to
the Johnson system
 \footnote [2]{http://heasarc.nasa.gov/W3Browse/all/tycho2.html};
 the
absolute magnitude $M_V$ was calculated without any extinction
correction. Main-sequence stars separated by the boundary
$B-V=0.^m6$ belong to groups I and II, group III includes giants,
and group IV consists of main-sequence dwarfs with an admixture of
subdwarfs. All of the stars used to construct the figure and to
determine the kinematic parameters for Table 2 have small errors
in the photometric data, more specifically, $\sigma_{B_T}<0.3^m$
and $\sigma_{V_T}<0.3^m$. As can be seen from the last row in the
table, the distance scale correction factor p is close to unity.
This means that there is no need to correct the scale of Gaia TGAS
trigonometric parallaxes.

Note the result obtained from the proper motions $(V_l,V_b)$ of
all stars from the Gaia TGAS catalogue with relative parallax
errors $\sigma_\pi/\pi<10\%$:
\begin{equation}
 \begin{array}{ccc}
 (U,V,W)_\odot= (9.28,20.35,7.36)\pm(0.05,0.07,0.05)~{\hbox {\rm km s$^{-1}$}},\\
     \Omega_0=27.24\pm0.30~{\hbox {\rm km s$^{-1}$ kpc$^{-1}$}},\\
 \Omega_0^{'}=-3.77\pm0.06~{\hbox {\rm km s$^{-1}$ kpc$^{-2}$.}}
 \end{array}
 \end{equation}
Here, the mean distance of the sample stars is $\overline
r=226$~pc, the number of stars used is $N_\star=424251$, the
linear circular rotation velocity at the Galactocentric distance
of the Sun is $V_0=218\pm6$ km s$^{-1}$ (for the adopted distance
$R_0=8.0\pm0.2$~kpc), and the Oort constants are
$A=15.07\pm0.25$~km s$^{-1}$ kpc$^{-1}$ and $B=-12.17\pm0.39$~km
s$^{-1}$ kpc$^{-1}$, $p=0.98\pm0.08$.

To determine the Galactic rotation parameters, of course, it is
more interesting to use stars at great distances. To use the data
on stars with relative parallax errors more than 10\%, we took
their distances from Astraatmadja and Bailer-Jones (2016b). These
distances were calculated using the trigonometric parallaxes of
stars from the Gaia TGAS catalogue and contain the corrections for
the Lutz–Kelker bias. We use the corrections calculated for two
stellar density distributions: (a) an exponential drop (away from
the Sun) with a radial scale length of 110 pc and (b) an
anisotropic distribution (dependent on the heliocentric distance
$r$ and coordinates $l, b$) close to the Milky Way model. For case
(b) the model parameters were selected (Astraatmadja and
Bailer-Jones 2016a) not from the spatial distribution of stars but
from the observed photometric characteristics (typical for Gaia
stars) by taking into account the peculiarities of the
distribution of absorbing matter in the Galaxy. We will call case
(b) the Milky Way model for short.

Tables 3 and 4 give the Galactic rotation parameters found only
from the proper motions of stars from the Gaia TGAS catalogue
using the calculated distances corrected for the Lutz–Kelker bias
for cases (a) and (b), respectively. When producing our samples,
we calculated the ratio $\sigma_\pi/\pi$ from the trigonometric
measurements; therefore, the corresponding columns in Tables 3 and
4 give the parameters derived from the same stars. When solving
Eqs. (2)--(3), we used the constraints on the proper motions
specified in (6).

 \begin{table}[t]
 \caption[]
  {\small
The Galactic rotation parameters found only from the proper
motions of stars $(V_l,V_b)$ from the Gaia TGAS catalogue using
the calculated distances from Astraatmadja and Bailer-Jones (2016)
corrected for the Lutz–Kelker bias for an exponential drop in
stellar density
  }
  \begin{center}  \label{t:03}
  \small
  \begin{tabular}{|l|r|r|r|r|r|}
   \hline
  Parameters           & $\sigma_\pi/\pi<0.15$ & $\sigma_\pi/\pi<0.20$ & $\sigma_\pi/\pi<0.30$ & $\sigma_\pi/\pi<0.50$ \\\hline

   $U_\odot,$    km s$^{-1}$    &  $ 9.10\pm0.04$   &  $ 8.90\pm0.03$   &  $ 8.57\pm0.03$   &  $ 8.25\pm0.02$  \\
   $V_\odot,$    km s$^{-1}$    &  $20.01\pm0.05$   &  $19.58\pm0.04$   &  $19.02\pm0.03$   &  $18.62\pm0.03$  \\
   $W_\odot,$    km s$^{-1}$    &  $ 7.15\pm0.03$   &  $ 6.90\pm0.03$   &  $ 6.53\pm0.02$   &  $ 6.21\pm0.02$  \\
 $\Omega_0,$     km s$^{-1}$ kpc$^{-1}$      &  $27.83\pm0.16$   &  $27.77\pm0.11$   &  $27.72\pm0.08$   &  $27.15\pm0.06$  \\
 $\Omega_0^{'},$ km s$^{-1}$ kpc$^{-2}$ &  $-3.835\pm0.035$ &  $-3.799\pm0.024$ &  $-3.788\pm0.017$ &  $-3.696\pm0.013$ \\
$\Omega_0^{''},$ km s$^{-1}$ kpc$^{-3}$ &  $ 5.789\pm0.168$ &  $ 4.397\pm0.095$ &  $ 3.298\pm0.054$ &  $ 2.803\pm0.037$ \\
   $\sigma_0,$   km s$^{-1}$    &           25.38   &           24.78   &          23.96    &           23.23  \\
               $A,$ km s$^{-1}$ kpc$^{-1}$  & $ 15.34\pm0.14$   & $ 15.20\pm0.10$   & $ 15.15\pm0.07$   & $ 14.78\pm0.05$  \\
               $B,$ km s$^{-1}$ kpc$^{-1}$   & $-12.48\pm0.21$   & $-12.58\pm0.15$   & $-12.57\pm0.10$   & $-12.36\pm0.08$  \\
             $\overline r,$ pc  &             297   &             348   &            410    &            461   \\
               $N_\star$        &          773520   &         1036878   &        1386239    &        1708304   \\
               $p$              &   $0.97\pm0.09$   &   $0.97\pm0.09$   &   $0.98\pm0.10$   &   $1.00\pm0.10$  \\
 \hline
 \end{tabular}\end{center} \end{table}
 \begin{table}[t]
 \caption[]
  {\small
The Galactic rotation parameters found only from the proper
motions of stars $(V_l,V_b)$ from the Gaia TGAS catalogue using
the calculated distances from Astraatmadja and Bailer-Jones (2016)
corrected for the Lutz–Kelker bias for the Milky Way model
  }
  \begin{center}  \label{t:04}
  \small
  \begin{tabular}{|l|r|r|r|r|r|}
   \hline
 Parameters           & $\sigma_\pi/\pi<0.15$ & $\sigma_\pi/\pi<0.20$ & $\sigma_\pi/\pi<0.30$ & $\sigma_\pi/\pi<0.50$ \\\hline

   $U_\odot,$    km s$^{-1}$    &  $ 9.23\pm0.04$   &  $ 9.14\pm0.03$   &  $ 9.03\pm0.03$   &  $ 9.02\pm0.02$  \\
   $V_\odot,$    km s$^{-1}$    &  $20.07\pm0.05$   &  $19.64\pm0.04$   &  $19.13\pm0.03$   &  $18.87\pm0.03$  \\
   $W_\odot,$    km s$^{-1}$    &  $ 7.26\pm0.03$   &  $ 7.11\pm0.03$   &  $ 6.96\pm0.02$   &  $ 6.97\pm0.02$  \\
 $\Omega_0,$     km s$^{-1}$ kpc$^{-1}$  &  $27.85\pm0.16$   &  $27.83\pm0.11$   &  $27.91\pm0.07$   &  $27.46\pm0.05$  \\
 $\Omega_0^{'},$ km s$^{-1}$ kpc$^{-2}$ &  $-3.819\pm0.034$ &  $-3.769\pm0.024$ &  $-3.766\pm0.016$ &  $-3.694\pm0.012$ \\
$\Omega_0^{''},$ km s$^{-1}$ kpc$^{-3}$ &  $ 4.614\pm0.156$ &  $ 2.798\pm0.081$ &  $ 1.425\pm0.039$ &  $ 0.864\pm0.021$ \\
   $\sigma_0,$   km s$^{-1}$    &           25.72   &           25.36   &          25.05    &           25.07  \\
               $A,$ km s$^{-1}$ kpc$^{-1}$  & $ 15.27\pm0.14$   & $ 15.08\pm0.09$   & $ 15.06\pm0.06$   & $ 14.78\pm0.05$  \\
               $B,$ km s$^{-1}$ kpc$^{-1}$   & $-12.58\pm0.21$   & $-12.75\pm0.14$   & $-12.85\pm0.10$   & $-12.68\pm0.07$  \\
             $\overline r,$ pc  &             303   &             363   &            446    &            537   \\
               $N_\star$        &          773466   &         1035443   &        1378648    &        1687383   \\
               $p$              &   $0.97\pm0.10$   &   $0.98\pm0.10$   &   $0.98\pm0.10$   &  $1.00\pm0.10$   \\
 \hline
 \end{tabular}\end{center}
 \end{table}

 \subsection*{DISCUSSION}
It is interesting to compare the Galactic rotation parameters
found in this paper with the results obtained in other papers. For
example, Bobylev and Bajkova (2016) analyzed stars with measured
line-of-sight velocities from the RAVE4 catalogue and proper
motions from the UCAC4 catalogue. The following kinematic
parameters were found from a sample of more than 145 000 stars:
 $(U,V,W)_\odot=(9.1,20.8,7.7)\pm(0.1,0.1,0.1)$~km s$^{-1}$,
 $\Omega_0=28.7\pm0.6$~km s$^{-1}$ kpc$^{-1}$, and
 $\Omega_0^{'}=-4.3\pm0.1$ km s$^{-1}$ kpc$^{-2}$, where
  $V_0=230\pm12$ km s$^{-1}$ (for the adopted distance $R_0=8.0\pm0.4$~kpc), as well
as the Oort constants
 $A=17.1\pm0.5$~km s$^{-1}$ kpc$^{-1}$ and
 $B=-11.6\pm0.8$~km s$^{-1}$ kpc$^{-1}$.

Rastorguev et al. (2017) determined the Galactic rotation
parameters from 136 masers with measured trigonometric parallaxes.
They used the VLBI measurements of water and methanol masers at
frequencies from 6 to 22 GHz that were performed by several
scientific teams using radio interferometers in the USA, Japan,
Europe, and Australia. These sources cover a wide range of
Galactocentric distances $R:0-16$~kpc. For example, for the C1
model (the model of a constant radial velocity dispersion) they
found
 $(U,V,W)_\odot=(11.0,19.6,8.9)\pm(1.4,1.2,1.1)$~km s$^{-1}$,
  $\Omega_0=28.4\pm0.5$~km s$^{-1}$ kpc$^{-1}$,
  $\Omega_0^{'}=-3.83\pm0.08$~km s$^{-1}$ kpc$^{-2}$,
  $\Omega_0^{''}=1.17\pm0.05$~km s$^{-1}$ kpc$^{-3}$, and
  $V_0=235\pm7$~km s$^{-1}$ (for the adopted
  $R_0=8.27\pm0.13$~kpc).

From the velocities of 260 Cepheids with measured proper motions
from the Gaia DR1 catalogue Bobylev (2017) found
 $(U,V,W)_\odot=(7.9, 11.7, 7.4)\pm(0.7, 0.8, 0.6)$ km s$^{-1}$,
  $\Omega_0 =28.8\pm0.3$~km s$^{-1}$ kpc$^{-1}$,
  $\Omega_0^{'}=-4.1\pm0.1$~km s$^{-1}$ kpc$^{-2}$, and
  $\Omega_0^{''}=0.81\pm0.07$~km s$^{-1}$ kpc$^{-3}$ (for the adopted
  $R_0=8.0\pm0.2$~kpc), the circular velocity
 $V_0=231\pm6$~km s$^{-1}$, as well as
 $A=16.2\pm0.4$~km s$^{-1}$ kpc$^{-1}$ and
 $B=-12.6\pm0.5$~km s$^{-1}$ kpc$^{-1}$.

From 238 OB stars using the proper motions from the Gaia DR1
catalogue, Bobylev and Bajkova (2017) found
 $(U,V,W)_\odot=(8.2,9.3,8.8)\pm(0.7,0.9,0.7)$~km s$^{-1}$,
 $\Omega_0=31.5\pm0.5$~km s$^{-1}$ kpc$^{-1}$,
 $\Omega_0^{'}=-4.4\pm0.1$~km s$^{-1}$ kpc$^{-2}$,
 $\Omega_0^{''}=0.71\pm0.10$~km s$^{-1}$ kpc$^{-3}$,
the Oort constants are
 $A=17.8\pm0.5$~km s$^{-1}$ kpc$^{-1}$ and
 $B=-13.8\pm0.7$~km s$^{-1}$ kpc$^{-1}$, and
 $V_0=252\pm8$~km s$^{-1}$ (for the adopted
 $R_0=8.0\pm0.2$~kpc).

The circular rotation velocity at the Galactocentric distance of
the Sun $V_0=218\pm6$~km s$^{-1}$ found in the solution (7) is
typically 15--25~km s$^{-1}$ smaller than the above values
obtained from samples of young stars. Such a discrepancy is
primarily attributable to the manifestation of Str\"omberg's
asymmetry effect (an increase in the lag of the mean rotation
velocity of a population of stars with increasing velocity
dispersion of this population). Since we use stars of all ages, we
obtain a reduced velocity $V_0.$

Using only the proper motions of $\sim$300 000 nearby ($<$250~pc)
main-sequence stars from the Gaia DR1 catalogue, Bovy (2017)
estimated the Oort constants $A, B, C,$ and $K$ that describe the
peculiarities of the local kinematics based on the Oort--Lindblad
model. In particular, he obtained the following estimates:
 $\Omega_0=27.1\pm0.5$~km s$^{-1}$ kpc$^{-1}$,
 $A=15.3\pm0.5$~km s$^{-1}$ kpc$^{-1}$ and
 $A=15.3\pm0.5$~km s$^{-1}$ kpc$^{-1}$, and
 $V_0=219\pm4$~km s$^{-1}$. It can be
estimated that the distance $R_0=8.1$~kpc was used here, while
 $\Omega_0^{'}$ (see Eq. (5)) is $-3.8$~km s$^{-1}$ kpc$^{-2}$. Note that our estimates of
the kinematic parameters in the solution (7), first, are as
accurate as the estimates of Bovy (2017) and, second, are more
reliable methodologically. Indeed, the Taylor expansion is used
twice in the Oort–Lindblad model that was used by Bovy. In the
first case, just as we did, the angular velocity $\Omega_0$ is
expanded into a series to terms of the first order of smallness in
$r/R_0$ and then the distance $R$ is expanded into a series in
powers of $r/R_0$ based on Eq. (4). In our approach the distance
$R$ is calculated from the exact formula (4) using the measured
trigonometric parallaxes of stars. Note that in our case (in
contrast to the Oort–Lindblad approach), the distance errors
$\sigma_{R_0}$ should be taken into account when calculating the
errors in the Oort constants $A$ and $B,$ as follows from Eqs.
(5). In spite of this, in the solution (7) we obtained smaller
errors in the Oort constants $A$ and $B$ than their errors derived
by Bovy (2017).

In this paper we considered the samples of stars that are not very
far from the Sun and, therefore, the second derivative of the
angular velocity of Galactic rotation $\Omega_0^{''}$ is
determined very poorly. As can be seen from the results presented
in Tables 3 and 4, the behavior of $\Omega_0^{''}$ in the second
case (Table 4) is most logical: they tend to $\approx0.8$~km
s$^{-1}$ kpc$^{-3}$ known from the analysis of completely
different data (for example, those listed at the beginning of this
section) with increasing radius of the sample. There are no other
significant differences between the parameters corresponding to
one another presented in Tables 3 and 4. Note that having compared
the distances derived by them for Cepheids, Astraatmadja and
Bailer-Jones (2016b) concluded that case (b) (the Milky Way model)
has an advantage for stars from the range of heliocentric
distances less than 2 kpc.

In this paper we applied a method that allows a ``multiplicative''
systematic error in the parallaxes to be detected. The
``additive'' error due to the error in the parallax zero point is
of great interest. To estimate this error, we found $\Omega_0^{'}$
for subsamples of stars with parallaxes in five different ranges.
The results are reflected in Table 5. However, the
``multiplicator'' $p$ turned out to have no significant dependence
on the mean parallax $\overline \pi$. In other words, based on the
data from the last two rows in Table 5, we attempted to determine
the correction $\Delta\pi$ from a linear equation like
$\overline\pi=A\cdot p+\Delta\pi$, but it turned out to be zero.

 \begin{table}[t]
 \caption[]
  {\small
The values of $\Omega_0^{'}$ found from the proper motions of
stars from the TGAS catalogue using the calculated distances from
Astraatmadja and Bailer-Jones (2017) corrected for the Lutz–Kelker
bias for the Milky Way model in five distance ranges  }
  \begin{center}  \label{t:05}
  \small
  \begin{tabular}{|l|r|r|r|r|r|r|}
   \hline
 $\overline r$    &         140 pc   &          251 pc  &         349 pc   &          448 pc   &          609 pc  \\\hline
 $\Omega_0^{'}$   & $-3.841\pm0.160$ & $-3.865\pm0.085$ & $-3.871\pm0.060$ &  $-3.928\pm0.050$ &  $-3.928\pm0.050$\\
 $N_\star$        &         193692   &         222917   &        215426    &         176114    &           227294 \\
 $p$              &  $0.96\pm0.09$   &  $0.96\pm0.07$   &  $0.96\pm0.08$   &  $0.94\pm0.08$    &     $1.00\pm0.09$\\
 $\overline \pi$  &         7.14 mas &         3.98 mas &         2.87 mas &        2.23 mas   &       1.64 mas   \\\hline
 \end{tabular}\end{center}
 {\small We used stars with $\sigma_\pi/\pi<0.20$,
 the values of $\Omega_0^{'}$ are given in km s$^{-1}$ kpc$^{-2}$.
 }
 \end{table}

We conclude that using the stellar proper motions from the Gaia
TGAS catalogue allows us to estimate the kinematic parameters of
our model in good agreement with the results of our analysis of
independent data. In this case, the distance scale of the Gaia
TGAS catalogue does not require using any additional correction
factor in the range of heliocentric distances less than
$\approx$1.5~kpc under consideration.

 \subsection*{CONCLUSIONS}
We considered the space velocities of stars calculated using their
highly accurate proper motions and trigonometric parallaxes from
the Gaia TGAS catalogue in combination with their line-of-sight
velocities from the RAVE5 catalogue. We obtained both simultaneous
and separate solutions of the basic kinematic equations, i.e., we
considered the equations that were set up using either only the
line-of-sight velocities of stars, or only their proper motions,
or a combination of all velocities. This allowed us to trace the
consistency between the data from a kinematic viewpoint.

Based on a sample of 53 173 Gaia--RAVE stars with relative
trigonometric parallax errors $\sigma_\pi/\pi$ less than 10\%, we
found the following kinematic parameters by simultaneously solving
the equations: the components of the group velocity vector for the
sample stars relative to the Sun
 $(U,V,W)_\odot=(9.15,20.41,7.74)\pm(0.18,0.15,0.12)$~km s$^{-1}$, the
angular velocity of Galactic rotation
 $\Omega_0=27.62\pm1.40$~km s$^{-1}$ kpc$^{-1}$, and its first derivative
 $\Omega_0^{'}=-3.89\pm0.23$ km s$^{-1}$ kpc$^{-2}$.
By comparing the values of $\Omega_0^{'}$ found by separately
analyzing the line-of-sight velocities
 ($\Omega_0^{'}=-3.70\pm0.38$~km s$^{-1}$ kpc$^{-2}$) and
proper motions
 ($\Omega_0^{'}=-3.83\pm0.29$~km s$^{-1}$ kpc$^{-2}$)
under the same constraints on the parallax error, we determined
the distance scale correction factor p to be close to unity,
$p=0.97\pm0.04.$

In the second part of the paper, we analyzed stars only from the
Gaia TGAS catalogue. Thus, we considered stars covering the entire
celestial sphere. From the proper motions of 424 250 stars with
relative trigonometric parallax errors less than 10\% we found
  $(U,V,W)_\odot=(9.28,20.35,7.36)\pm(0.05,0.07,0.05)$~km s$^{-1}$,
  $\Omega_0=27.24\pm0.30$~km s$^{-1}$ kpc$^{-1}$, and
  $\Omega_0^{'}=-3.77\pm0.06$~km s$^{-1}$ kpc$^{-2}$;
here, the circular rotation velocity of the Sun around the
Galactic center is
 $V_0=218\pm6$ km s$^{-1}$ (for the adopted distance
 $R_0= 8.0\pm0.2$ kpc) and the Oort constants are
 $A= 15.07\pm0.25$~km s$^{-1}$ kpc$^{-1}$ and
 $B=-12.17\pm0.39$~km s$^{-1}$ kpc$^{-1}$,
  $p=0.98\pm0.08.$ All
of the parameters obtained in this solution are new, among the
most reliable ones to date.

The sample of 424 250 Gaia TGAS stars was divided into four
subgroups: giants (I), main-sequence dwarfs (II), giants (III),
and subdwarfs (IV). A comparison of $\Omega_0^{'}$ found by
analyzing the proper motions of the stars from these subgroups
with its value found previously only from the line-of-sight
velocities of stars from the RAVE5 catalogue showed the distance
scale correction factor to be $p>0.93.$

The kinematics of more distant stars (with relative parallax
errors more than 10\%) was studied by invoking the distances from
Astraatmadja and Bailer-Jones (2016), which were calculated using
the trigonometric parallaxes of stars from the Gaia TGAS catalogue
and contain the corrections for the Lutz-Kelker bias. We
considered the distances to which the corrections were applied for
two stellar density distributions: (a) an exponential drop with a
radial scale length of 110~pc and (b) an anisotropic distribution
close to the real Milky Way model. For both stellar density
distributions we determined the kinematic parameters for samples
with various relative distance errors (15\%, 20\%, 30\%, 50\%) and
found the distance scale correction factor p to be always close to
unity. Finally, we showed that at mean parallax errors for the
sample $\sigma_\pi/\pi<50\%$ more reliable kinematic parameters
are obtained for case (b). For example, in contrast to case (a),
at a small mean stellar distance $\overline{r}=537$~pc the
calculated second derivative of the angular velocity of Galactic
rotation$\Omega_0^{''}=0.864\pm0.021$ km s$^{-1}$ kpc$^{-3}$ is in
good agreement with the results of other authors.

On the whole, we found that the distances to the stars (both
giants and dwarfs) from the Gaia TGAS catalogue that were
calculated using their trigonometric parallaxes do not require
using any additional correction factor.

 \subsubsection*{ACKNOWLEDGMENTS}
We are grateful to the referee for the useful remarks that
contributed to an improvement of the paper.  This work was
supported by the Basic Research Program P--7 of the Presidium of
the Russian Academy of Sciences, the ``Transitional and Explosive
Processes in Astrophysics'' Subprogram.

 \bigskip\medskip{\bf REFERENCES}
{\small

1. D. An, D. M. Terndrup, M. H. Pinsonneault, D. B. Paulson, R. B.
Hanson, and J. R. Stauffer, Astrophys. J. 655, 233 (2007).

 2. T. L. Astraatmadja and C. A. L. Bailer-Jones, Astrophys. J. 832, 137 (2016a).

3. T. L. Astraatmadja and C. A. L. Bailer-Jones, Astrophys. J.
833, 119 (2016b).

4. G. F. Benedict, B. E. McArthur, E. P. Nelan, and T. E.
Harrison, Publ. Astron. Soc. Pacif. 129, 2001 (2017).

5. V. V. Bobylev and A. T. Bajkova, Astron. Lett. 42, 90 (2016).

6. V. V. Bobylev, Astron. Lett. 43, 152 (2017).

7. V. V. Bobylev and A. T. Bajkova, Astron. Lett. 43, 159 (2017).

8. J. Bovy, Mon. Not. R. Astron. Soc. 468, L63 (2017).

9. A. G. A. Brown, A. Vallenari, T. Prusti, J. de Bruijne, F.
Mignard, R. Drimmel, et al. (GAIA Collab.), Astron. Astrophys.
595, A2 (2016).

10. S. Casertano, A. G. Riess, B. Bucciarelli, and M. G. Lattanzi,
Astron. Astrophys. 599, 67 (2017).

11. G. Clementini, L. Eyer, V. Ripepi, M. Marconi, T. Muraveva, A.
Garofalo, L. M. Sarro, M. Palmer, et al., Astron. Astrophys. 605,
79 (2017).

12. E. Hog, C. Fabricius, V. V. Makarov, U. Bastian, P.
Schwekendiek, A. Wicenec, S. Urban, T. Corbin, and G. Wycoff,
Astron. Astrophys. 355, L27 (2000).

13. W.-C. Jao, T. J. Henry, A. R. Riedel, J. G.Winters, K. J.
Slatten, and D. R. Gies, Astrophys. J. 832, L18 (2017).

14. B. Kim, D. An, J. R. Stauffer, Y. S. Lee, D. M. Terndrup, and
J. A. Johnson, Astrophys. J. Suppl. Ser. 222, 19 (2016).

15. A. Kunder, G. Kordopatis, M. Steinmetz, T. Zwitter, P.
McMillan, L. Casagrande, H. Enke, J. Wojno, et al., Astron. J.
153, 75 (2017).

16. F. van Leeuwen, Astron. Astrophys. 497, 209 (2009).

17. F. van Leeuwen, A. Vallenari, C. Jordi, L. Lindegren, U.
Bastian, T. Prusti, J. H. J. de Bruijne, A. G. A. Brown, C.
Babusiaux, et al. (GAIA Collab.), Astron. Astrophys. 601, 19
(2017).

18. L. Lindegren, U. Lammers, U. Bastian, J. Hernandez, S.
Klioner, D. Hobbs, A. Bombrun, D. Michalik, et al., Astron.
Astrophys. 595, A4 (2016).

19. A.-Li Luo, Y.-H. Zhao, G. Zhao, Li-C. Deng, X.- W. Liu, Yi-P.
Jing, G. Wang, H.-T. Zhang, et al., Res. Astron. Astrophys. 15,
1095 (2015).

20. T. E. Lutz and D. H. Kelker, Publ. Astron. Soc. Pacif. 85, 573
(1973).

21. C. Melis, M. J. Reid, A. J. Mioduszewski, J. R. Stauffer, and
G. C. Bower, Science 345, 1029 (2014).

22. T. Prusti, J.H. J. de Bruijne, A. G. A. Brown, A. Vallenari,
C. Babusiaux, C. A. L. Bailer-Jones, U. Bastian, M. Biermann, et
al. (GAIA Collab.), Astron. Astrophys. 595, A1 (2016).

23. A. S. Rastorguev, M. V. Zabolotskikh, A. K. Dambis, N. D.
Utkin, A. T. Bajkova, and V. V. Bobylev, Astrophys. Bull. 72, 122
(2017).

24. R. Sch\"onrich and M. Aumer, MNRAS 472, 3979 (2017).

25. K. G. Stassun and G. Torres, Astrophys. J. 831, 74 (2016).

 26. M. Steinmetz, T. Zwitter, A. Siebert, F. G. Watson, K. C. Freeman, U. Munari,
     R. Campbell, M. Williams, et al., Astron. J. 132, 1645 (2006).

 27. J. P. Vall\'ee, Astrophys. Space Sci. 362, 79 (2017).

 28. V. V. Vityazev, A. S. Tsvetkov, V. V. Bobylev, and A. T. Bajkova,
     Astrophysics 60, 462 (2017).

 29. M. V. Zabolotskikh, A. S. Rastorguev, and A. K. Dambis, Astron. Lett. 28, 454 (2002).

 30. The Hipparcos and Tycho Catalogues, ESA SP--1200 (1997).

 }

\end{document}